\documentstyle[epsfig]{aipproc}

\begin{document}
\title{Light Cone Consistency in Bimetric General Relativity}

\author{J. Brian Pitts$^*$ and W.C. Schieve$^*$}
\address{$^*$The Ilya Prigogine Center for Studies in Statistical Mechanics and Complex
Systems
\\Department of Physics      
\\
The University of Texas at Austin
\\
Austin Texas 78712}

\maketitle

\begin{abstract}
General relativity can be formally derived as a flat spacetime theory, but the consistency of the
resulting curved metric's light cone with the flat metric's null cone has not been adequately
considered.  If the two are inconsistent, 
then gravity is not just another field in flat spacetime after all. 
 Here we discuss recent progress in describing the conditions for consistency and prospects for
satisfying those conditions.
\end{abstract}

\section*{Introduction}
The formulation and derivation of general relativity using a flat metric tensor $\eta_{\mu\nu}$
are well-known from the works of Rosen, Gupta, Kraichnan, Feynman,
 Deser,  Weinberg \emph{et al.}
\cite{SliBimGRG}.  One
can obtain a curved metric $g_{\mu\nu}$ by adding the gravitational potential
$ \gamma_{\mu\nu}$ to the flat metric $\eta_{\mu\nu}$:
\begin{eqnarray}
g_{\mu\nu} = \eta_{\mu\nu} + \sqrt{32 \pi G} \gamma_{\mu\nu}.
\end{eqnarray}
This framework is useful \cite{PetrovNarlikar}, but is it merely formal?
If general relativity can be \emph{consistently}
 regarded as  a special-relativistic theory, then the observable curved metric must satisfy
a nontrivial consistency condition in relation to 
the unobservable flat background metric:
 the ``causality principle'' says that the curved metric's light cone cannot
open wider than the flat metric's.  The question of the relation between the cones is complicated
somewhat by its gauge-variance. 

\section*{Previous Discussions of Relationship of Null Cones}

     While the flat spacetime field approach to general relativity has been mature since the
1950s, the question of the consistency of the effective curved metric's null
 cone with the original flat metric's received surprisingly little attention. In the 1970s van
Nieuwenhuizen wrote: ``The strategy of
particle physicists has been to ignore [this problem] for the time being, in the hope that [it] will
ultimately be resolved in the final theory.  Consequently we will not discuss [it] any further.'' 
\cite{van N cones}
More recently (since the late 1970s), this issue has received more sustained attention
\cite{Penrose,Zel2,BurlankovCause,LogunovBasic}, but the treatments to date have 
been impaired by unnecessarily strict requirements \cite{Penrose,BurlankovCause} or lack of a
general and systematic approach  \cite{Zel2,LogunovBasic}, as we have noted
\cite{SliBimGRG}.  

     We propose to \emph{stipulate} that the gauge be fixed in a way that the proper relation
obtains, if possible.  The gauge fixing can be implemented in an action principle using ineffective constraints, whose constraint forces vanish 
\cite{PSShepley}.  
This approach does appear to be possible, because the gauge freedom allows one to choose arbitrarily
$g^{00}$ and $g^{0i}$ (at least locally).  Increasing $g^{00}$ stretches the curved metric's null
 cone along the time axis, so that it becomes narrower, while adjusting $g^{0i}$ controls the tilt
of the curved null cone relative to the flat one.  Stretching alone appears to be enough to satisfy the causality principle,
in fact. 

\section*{Kinematic and Dynamic Progress}
     
     The metric is a poor variable choice due to the many off-diagonal 
terms.  One would like to diagonalize
$g_{\mu\nu}$ and $\eta_{\mu\nu}$ simultaneously by solving the generalized eigenvalue
problem $ g_{\mu\nu} V^{\mu} = \Lambda \eta_{\mu\nu} V^{\mu},$ but in general that is
impossible, because there is not a complete set
of eigenvectors on account of the minus sign in $\eta_{\mu\nu}$ \cite{HallNegm}.
There are 4
Segr\'{e} types for a real symmetric rank 2 tensor with respect to a Lorentzian metric,
the several types having different numbers and sorts of
eigenvectors \cite{HallNegm}. 
     We have recently used this technology to classify $g_{\mu\nu}$ with respect to 
$\eta_{\mu\nu}$.  Two types are forbidden by the causality principle.  One type has members that
obey the causality principle, but we argue that they can be ignored.  The remaining type has 4 real
independent orthogonal eigenvectors, as one would hope.  In that case, the causality principle is
just the requirement that the temporal eigenvalue be no larger than each of the three spatial
eigenvalues. 

\begin{figure}[b!] 
\centerline{\epsfig{file=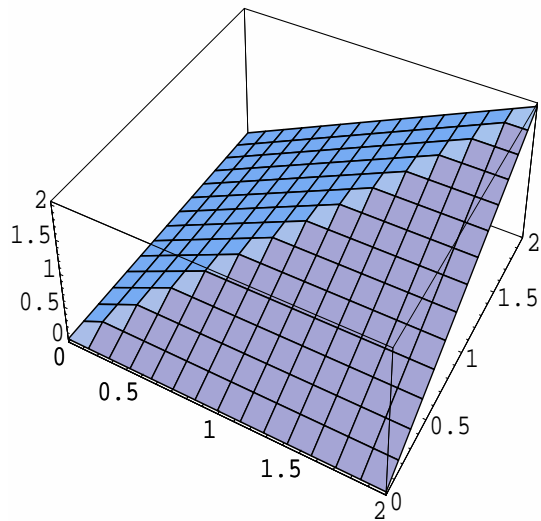,height=3.5in,width=3.5in}}
\vspace{10pt}
\caption{Bounding Surface for Temporal Eigenvalue as Function of Spatial Eigenvalues in 2
Dimensions}
\label{fig}
\end{figure}

     Realizing the condition $g_{\mu\nu} \rightarrow  \eta_{\mu\nu}$ 
when the gravitational field is weak, while obeying the causality principle, is nontrivial. The
causality principle
puts an
upper 
bounding surface on the temporal eigenvalue in terms of the spatial ones, and the surface is
\emph{folded}, as seen in 2 spatial dimensions in figure \ref{fig}.
     Einstein's  equations have second spatial derivatives of
$g^{00}$ (which is closely related to the temporal eigenvalue), so the fold, if not avoided, would
imply Dirac
delta gravitational `forces' that make the canonical momenta jump discontinuously.  
On the other hand, avoiding the fold means excluding $g_{\mu\nu} =  \eta_{\mu\nu}$!   
 But why fix the temporal eigenvalue in terms of the
 spatial eigenvalues at the same point only (ultralocally)?  It is enough to do so locally, by
admitting derivatives.  
 When the derivatives are nonzero, the fold is avoided, but as they vanish, the fold is approached.
If such a partial gauge-fixing  can be found, then it will facilitate interpreting the Einstein equations as describing a special-relativsitic field theory.  In such a theory, 
one would need to consider the physical situation near the Schwarzschild radius rather carefully.


\end{document}